\documentstyle [12pt]{article}
\def\theta{\vartheta}

\begin{document}

\mbox{ }\\
\hspace*{110mm} MPI-Ph/93-40\\
\hspace*{110mm} Saclay T/93/063\\
\hspace*{110mm} June 1993

\vspace{10mm}
\centerline{\Large\bf Finite size scaling analysis}
\centerline{\Large\bf of intermittency moments}
\centerline{\Large\bf in the two dimensional Ising model }
\vspace*{1cm}
\begin{center}
{\large Z. Burda} $^{a}$, {\large and K. Zalewski} $^{b,c}$\\
\vspace*{1.5mm}
Institute of Physics, Jagellonian University\\
Reymonta 4, PL-30 059 Cracow, Poland \\
\mbox{ } \\
{\large R. Peschanski} \\
\vspace*{1.5mm}
Service de Physique Th\'{e}orique\\
Centre d'Etudes de Saclay, F-91191 Gif-sur-Yvette, Cedex, France\\
\mbox{ } \\
{\large J. Wosiek} $^{d}$ \\
\vspace*{1.5mm}
Max Planck Institut f\"ur Physik \\
W-8000 Munich 40, P.O. Box 401212, Germany \\
\end{center}
\vspace*{1cm}

\begin{abstract}
Finite size scaling is shown to work very well for the block
variables used in intermittency studies on a 2-d Ising lattice.
The intermittency exponents
so derived exhibit the expected relations to the magnetic critical
exponent of the model.
\end{abstract}

\vspace{\fill}
\noindent
{\small
$^a)$ Partly supported by the KBN grant 2P30216904. \newline
$^b)$ Partly supported by the KBN grant 203809101. \newline
$^c)$ Also Institute of Nuclear Physics, Cracow. \newline
$^d)$~On leave from the Jagellonian University, Cracow.
}
\newpage

{\em Introduction.} Intermittency studies have developed into
an active branch of particle physics. In spite of much effort, however,
the origin of intermittency in multiparticle production processes
is still controversial. In the original papers \cite{BP86} two possible
mechanisms have been pointed out. Intermittency could be: a result of
particle production by long random cascades, or a signal of a phase
transition. It was soon suggested (for early reference cf. \cite{KZ})
that it also could result from suitably conspiring ordinary processes
like resonance production and decay, Bose-Einstein interference etc.
Standard QCD cascade was also shown to lead to intermittency
at very high energies \cite{OW}. In order to shed some light
on possible mechanisms, it is instructive to consider solvable systems
where intermittency is present and its origin is unambiguous.

It has been pointed out by one of us \cite{W88} that intermittency
should be observable at the phase transition point of the two-dimensional
(2d) Ising model. Indeed the transition is of
second order, hence the fluctuations are selfsimilar at all length
scales.
This is usually the condition for the
 occurrence of intermittency. In order to define it quantitatively,
the $L\times L$ Ising lattice is divided into cells of size
$l\times l$. For each cell a block variable $k_l$ is defined, and
the thermodynamical averages of moments of $k_l$ are computed
using standard Monte Carlo methods. If the moments $M_p$ show a power
dependence on the size of a cell, i.e. if
\begin{equation}
\log{M_p} = -\lambda_p \log{l} +c_p ,     \label{F1}
\end{equation}
where $\lambda_p$ and $c_p$ do not depend on $l$, the system is
said to be intermittent with the intermittency exponents $\lambda_p$.

Such a formulation of the problem bears a striking resemblance to the
renormalization group (RG) approach used to study critical
phenomena \cite{RG}. Indeed, employing results of the RG analysis,
Satz
\cite{S89} derived a relation between intermittency exponents and
the known critical magnetic exponent $\delta$ of the Ising model.
Let us choose as the block variable
\begin{equation}
       k_l=|\overline{s}_l|,    \label{F2}
\end{equation}
where $\overline{s}_l={1\over l^2} \sum_{i\in C_l} s_i$
denotes the average spin in an $l\times l$ cell $C_l$. Satz used
$\overline{s}_l$ without the absolute value, which leads to some
difficulties, but the idea is the same. According to the RG \cite{MA},
when calculating thermodynamical averages,
the dynamical variables $\overline{s}_l$ can be replaced by
\begin{equation}
 \tilde{s}_l = l^x \overline{s}_l,
 \label{F2a}
\end{equation}
where the new variables $\tilde{s}_l=\pm 1$ just like the original spins.
The constant $x$ is known as the anomalous dimension of the spin field.
For the two-dimensional Ising model $x$ can be expressed by the critical
magnetic exponent $\delta=15$ (cf. \cite{S89}). One finds
\begin{equation}
x={2\over \delta+1}={1\over 8}
\label{F3}
\end{equation}
Using Eq.(\ref{F2a}) one obtains the power dependence on the block size
\begin{equation}
<|\overline{s}_l|^p>=l^{-p/8} g({l\over L},{L\over \xi(\beta)}).
\label{F4}
\end{equation}
Coefficient $g$ depends in general on all independent ratios of the
 dimensional parameters entering the problem. In particular the
 dependence
on temperature enters through the ratio of the infinite volume
correlation length $\xi(\beta)$ to any other relevant scale parameter.
Assuming that $g(0,L/\xi(\beta)) \neq 0 $, the above finite size scaling
Ansatz gives Eq.(\ref{F1}), for $1<<l<<L$ with the
intermittency exponents
\begin{equation}
\lambda_p = {1 \over 8  } p .   \label{F6}      \end{equation}
Let us stress that
Eq.(\ref{F4}) represents only the leading, in $l$,
 result of the RG analysis. In general nonleading terms may affect this
simple behaviour.

A convincing numerical confirmation of Eq.(\ref{F6})
was, however, more difficult. In Ref. \cite{W88} the
number of ``up" spins
in a cell $ {1\over 2}(1+\overline{s}_l) $  was chosen as a
block variable.
In the vicinity of the critical temperature the increase of
the normalized
moments with decreasing cell size was demonstrated qualitatively.
Later \cite{BFS90,GLS91} it was pointed out that $Z_2$ symmetric
observables may be more convenient
 for quantitative determination of the intermittency indices, since
they are insensitive to fluctuations between the two ordered phases. Such
transitions, which occur in finite systems, may mask the true critical
fluctuations.

The authors of Refs \cite{BFS90,GLS91} suggested the block variable
${1\over 2}(1+\overline{S}_l)$, where
\begin{equation}
\overline{S}_l=\overline{s}_l \; {\rm sign}(\overline{s}_L),  \label{F8}
\end{equation}
They also choose to work at the quasicritical temperature
where, at given $L$, the magnetic susceptibility is the largest.
At first they did not get a clear signal for the power dependence
\cite{BFS90}, but an improved analysis \cite{GLS91}
uncovered the selfsimilar behaviour;
however, rather surprisingly, with intermittency indices derived from the
percolation critical exponent and not from the magnetic one. A possible
explanation of this result was suggested in \cite{BWZ}. Moments of this
particular block variables are linear combinations of averages of
various powers of $\overline{S}_l$. Assuming for each power of
$\overline{S}_l$ a scaling law analogous to (\ref{F1}),
 one indeed finds, for the lattice sizes considered,
 the effective exponents
 similar to the percolation ones. Whether this is a numerical
 coincidence,
or evidence for a genuine relation to percolation, is an open problem
\cite{SV,P}. Very recently Leroyer \cite{L} used $\overline{S}_l$ as the
block variable and for $\beta=\beta_c={1\over 2}\log{(1+\sqrt{2})} $
got a clear confirmation of the
power behaviour with intermittency exponents derived from the
magnetic one. We shall comment on his results later.

{\em Results.}
The main points of the present work are: the choice of
$|\overline{s}_l|$
as the block variable, and a consistent finite size scaling (FSS)
analysis of our Monte Carlo data. In the FSS approach the problem of
choosing the best $\beta$ is less severe, since one is
studying the dependence
on $L$ and $\beta$ in a small scaling window around $\beta_c$.
In the 2d Ising model the critical exponent $\nu$ is equal to $1$, hence
the FSS variable is
\begin{equation}
{L\over \xi(\beta)} \simeq L \Delta \beta\equiv y, \;\;\;
\Delta \beta = \beta-\beta_c.
\label{F9}
\end{equation}
Our runs were done for four
values of $y$ $(0.0,0.105,0.21,0.42)$, i.e. in the range
which corresponds to $\beta$ between $\beta_c$ and the pseudocritical one
for given $L$. We find numerically that the results depend weakly on the
choice of $y$ in this region.
\begin{figure}[htb]
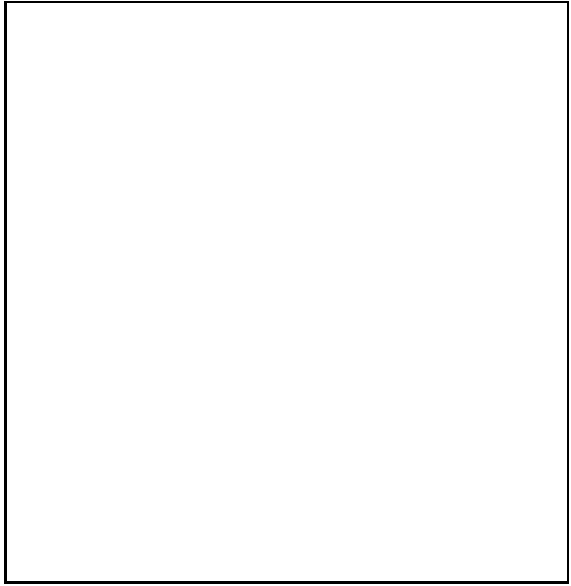

\centerline{\framebox[75mm]{\rule[-21mm]{0mm}{75mm}}}
\caption{Dependence of the second moment, for $y=0.105$, on the cell
size $l$. Continuous lines correspond to fixed ${l\over L}$ (see text).
The dashed line connects points for $L=1024$.}
\label{fig:f1}
\end{figure}

MC simulations were performed using the
Swendsen-Wang algorithm \cite{SW}.
The values $L=64,128,256,512,1024$ and $4\leq l \leq L$
were used. For each
lattice size about $10^5$ measurements were done. In order to reduce
autocorrelations a measurement was taken every sixth sweep,
for smaller lattices,
and every eight sweep for larger ones. We took advantage of the
very slow growth
of the autocorrelation time in the Swendsen-Wang method.
The critical exponent
characterizing this growth is estimated to be about  $.35$. For each new
lattice the first $10^4$ sweeps were discarded to allow
the system to reach
equilibrium. Statistical errors of our MC results are
always less than the sizes of the symbols used to denote points on the
figures.
At fixed $l/L$ and $y$, to a very good
accuracy, the dependences of
$\log <|\overline{s}_l|^p>$  on $\log l$ are found to be linear in all
the regions considered. The slopes agree with predictions
(\ref{F6}) within small
errors, which can be ascribed to the finite $L$ corrections.

In Fig.1 we show as an example the dependence of
$\log <\overline{s}_{l}^2>$
on $\log l$ for  fixed $y=0.105$ and fixed $l/L=2^{-m}, m=0,1,\dots ,6$.
The dotted line
joins the points with fixed $L=1024$. The usual
approach was to use such lines
for $l<<L$ to find the slopes. Clearly, the slopes of lines evaluated at
constant ${l\over L}$ are much better defined in
accordance with Eq.(\ref{F4}).
The slopes for the six lines in the figure range
from $0.239$ to $0.272$ to be
compared with ${1\over 4}$ obtained from the magnetic
exponent, Eq.(\ref{F6}),
and ${10 \over 96} \simeq .104$ resulting from the percolation
exponent.
Certainly the magnetic exponent is preferred. In Fig. 2  we show
the results
for $y=0.42, l/L=1/8$ and $p=1,2,3,4,5$. The lines are normalized
at $l=32$
and have slopes $p/8$. The agreement is quite satisfactory.
\begin{figure}[htb]
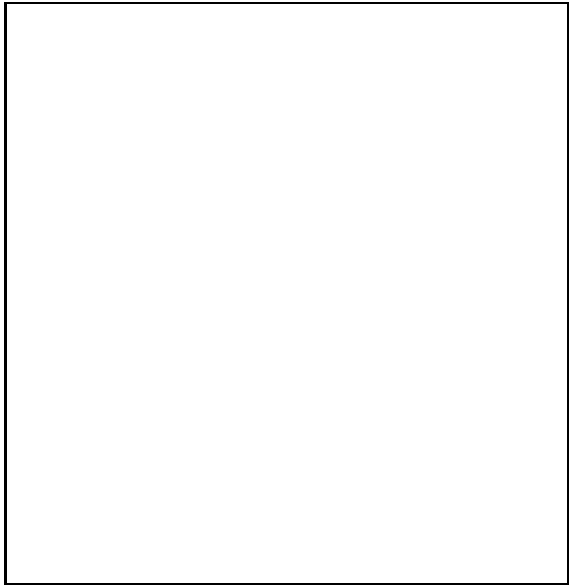

\centerline{\framebox[75mm]{\rule[-21mm]{0mm}{75mm}}}
\caption{Dependence of the moments $p=1,\dots ,5$ for $y=0.42$
on the cell
size $l$. The lines are calculated from ${\rm log}_2
M_p(l)={\rm log}_2
M_p(32)
-{p\over 8}
({\rm log}_2 l -5)$.}
\label{fig:f2}
\end{figure}

The normalized moments \cite{B} are of little interest for the present
choice of the
block variable, because from Eq.(\ref{F4})
\begin{equation}
F_p \equiv {<|\overline{s}_l|^p> \over <|\overline{s}_l|>^p } \sim l^0,
\;\;\;\;
l/L=const, \label{F12}
\end{equation}
where $\sim$ means equal up to a scale independent factor.
At fixed $L$ there
could be some $l$ dependence from the argument of the scale invariant
coefficients. However, since our numerical results show that
these coefficients
tend to nonzero limits
as $l/L \rightarrow 0$, there are no
relations analogous
to Eq.(\ref{F1}) with non-zero $\lambda_p$ for $l<<L$.

It is a property
of the Ising model, and in fact of many field theories,
that the first power
of a field has a nontrivial anomalous dimension.
In the multiparticle language single particle inclusive distributions
are not smooth.
The anomalous dimension of the $n$-th power of the field is
in our case simply $n$ times the anomalous dimension of the
field itself. Thus the anomalous dimensions in the numerator
and denominator of Eq.(\ref{F12}) cancel.
Fortunately in order to study
selfsimilarity or intermittency it is enough
to normalize
moments, Eq.(\ref{F4}),
by appropriate powers of the volume of a cell \cite{W89},
which looses no
information about critical indices.

There is also no motivation to use the factorial \cite{BP86} or
binomial \cite{N}
moments, since, although we always consider the $l<<L$ limit,
one should keep
$1 << l$ at the same time,  i.e. the ``number of particles" in a cell
must be large.
This means that the bias introduced by the noise is not important,
and
tricks invented to filter it out are not necessary.

We discuss now moments of the block variable $\overline{S}_l$,
Eq.(\ref{F8}).
For $p$ even and/or $l=L$ we have
$\overline{S}_{l}^p = |\overline{s}_l|^p$.
Thus  only the moments with $p$ odd and $l<L$ require a
separate discussion.
As we will see shortly, these moments have a different scaling
behaviour, which
is also confirmed by our data. The main difference with the
previous block
variable is that, due to the translational invariance and additivity,
 the first moment
of $\overline{S}_l$ is {\em independent} of the cell size
\footnote{One may say
that $\overline{S}_l$ has zero anomalous dimension, however
any higher power
does not.}. Indeed we have
\begin{equation}
<\overline{S}_l>=<<\overline{S}_l>>=<\overline{S}_L>\sim L^{-1/8}.
\label{F13a}
\end{equation}
Where the double brackets denote average over cells and over
the ensemble,
and $\sim$ means, as usual, equal up to a scale invariant coefficient.
Assuming that the RG result, Eq.(\ref{F2a}), holds for the new variable
\begin{equation}
<\overline{S}^{p}_l> \sim l^{-p/8}<\tilde{S}_l>,  \;\;\; p - odd,
   \label{F13}
\end{equation}
where we have used $\tilde{S}_{l}^p=\tilde{S}_l$ for odd $p$.
By definition we have
$\tilde{S}_L=|\tilde{s}_L|=1$.
Equations (\ref{F13a}) and (\ref{F13}) imply
\begin{equation}
<\tilde{S}_l> \sim \left ( {l\over L}\right )^{1/8},
\label{F14}
\end{equation}
while for the normalized moments, studied also by Leroyer \cite{L}
one gets
\begin{equation}
F_p \equiv {<\overline{S}_{l}^p> \over <\overline{S}_l>^p } =
\left ( {l \over L}\right )^{{1-p \over 8}} h({l\over L},y),
\;\;\; p - odd,
\label{F15}
\end{equation}
where the scale invariant function
has been written explicitly.
Again, for $h(z,y)$
regular and non-vanishing at $z=0$, the
above formula
gives the power behaviour with the intermittency exponents
\begin{equation}
\overline{\lambda}_p={p-1\over 8},  \;\;\; p - odd.    \label{F16}
\end{equation}
The behaviour (\ref{F15}) has been numerically confirmed by Ref.\cite{L}.
On the other hand for the unnormalized moments we get
\begin{equation}
<\overline{S}^{p}_l>=l^{-{p-1\over 8} } L^{-{1\over 8}}
H({l\over L},y).   \label{F17}
\end{equation}
At fixed $L>>l$ these
 moments scale as $<|\overline{s}_l|^{p-1}>$ while
for $l\approx L$ they coincide with $<|\overline{s}_l|^p>$. Both these
features are clearly seen in Fig.3, where $\log{<\overline{S}_{l}^p>}$
$(p=1,3,5)$ and $\log{<|\overline{s}_l|^p>}$ $(p=1,2,3,4,5)$ are plotted
versus $\log{l}$ at fixed $L=512$. Note that also $<\overline{S}_l>$
for $l<<L$ is parallel to $<|\overline{s}_l|^0>=1$.

\begin{figure}[htb]
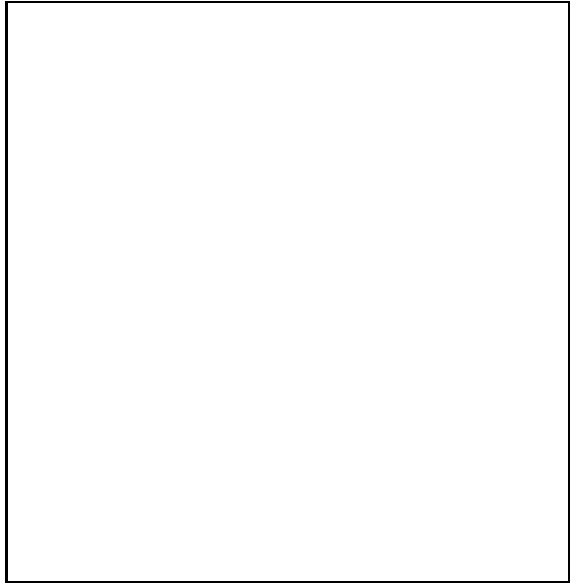

\centerline{\framebox[75mm]{\rule[-21mm]{0mm}{75mm}}}
\caption{Dependence of the moments, for $y=0.21$, on the cell
size $l$. The dashed lines correspond to the block variable
$|\overline{s}_l|$ for $p=1,2,3,4,5$ and solid lines to $\overline{S}_l$
for $p=1,3,5$. Note that the solid lines are parallel to the dashed
ones with $p\rightarrow p-1$. }
\label{fig.f3}
\end{figure}

{\em Summary.} It is seen that the intermittency of the Ising model
is yet another consequence of the well known scale invariance
of the system in the critical region.
The renormalization group formula, together with the finite size scaling
(\ref{F4}), give a good description of this phenomenon in the vicinity
of $\beta_c \;\;( 0\leq L \Delta \beta \leq .42 )$, for lattice sizes
$ L=64 $ to $1024$ and cell sizes $l=4$ to $L$. The power behaviour,
Eq.(\ref{F1}),
was confirmed with the intermittency exponents determined by the magnetic
critical exponent $\delta$ when $|\overline{s}_l|$
or $\overline{S}_l$ are used
as block variables. Moments of both variables were
shown to obey simple
relations which are also confirmed by our data. On the other hand,
it is still an open question
whether it is possible to find a block variable, which
would yield intermittency exponents related to the percolation exponent.
In particular the proposal \cite{P} to consider the largest connected
cluster
in a cell remains to be studied.

We would like to stress a few points which considerably simplify both
the theoretical and the numerical studies.  \newline
$\bullet$ The absolute value of the resultant spin of a cell
$|\overline{s}_l|$  is $Z_2$ symmetric and seems to be a convenient
choice for the block variable. \newline
$\bullet$ An advantage of the finite size scaling approach is
that instead
of selecting a particular value of $\beta$ one can
study the simultaneous dependence on $\beta$ and $L$ by fixing
$y=L\Delta\beta$ at several not too
large values. This is both simpler
and less restrictive. \newline
$\bullet$ The finite size scaling analysis at fixed $l/L$ is simpler
conceptually, and gives better defined slopes, than the analysis at fixed $L$.
This implies that (nearly) exact power behaviour (\ref{F1}) at fixed $L$
can be expected only at $l<<L$ (keeping still $1<<l$).

\vspace*{1cm}
K. Z. thanks for hospitality
 the Service de Physique
Th\' eorique de Saclay, where this work was started, and
the Max-Planck-Institute f\"{u}r Physik
in M\"{u}nchen, where it has been completed.

\end{document}